\documentclass{article}

\usepackage{microtype}
\usepackage{graphicx}
\usepackage{subfigure}
\usepackage{booktabs} 
\usepackage{multirow}
\usepackage{amsmath}
\usepackage{enumitem}

\usepackage{hyperref}

\usepackage[accepted]{mlsys2025}

\begin{document}

\twocolumn[
\mlsystitle{FlowMesh: A Service Fabric for Composable LLM Workflows}


\begin{mlsysauthorlist}
\mlsysauthor{Junyi Shen}{nus}
\mlsysauthor{Noppanat Wadlom}{nus}
\mlsysauthor{Lingfeng Zhou}{sjtu}
\mlsysauthor{Dequan Wang}{sjtu}
\mlsysauthor{Xu Miao}{dc}
\mlsysauthor{Lei Fang}{dc}
\mlsysauthor{Yao Lu}{nus}
\end{mlsysauthorlist}

\mlsysaffiliation{nus}{National University of Singapore}
\mlsysaffiliation{dc}{DataCanvas}
\mlsysaffiliation{sjtu}{Shanghai Jiao Tong University}

\mlsyscorrespondingauthor{Yao Lu}{yao@nus.edu.sg}

\mlsyskeywords{Machine Learning, MLSys}

\vskip 0.3in

\begin{abstract}
AI deployment increasingly resembles a pipeline of data transformation, fine-tuning, and agent interactions rather than a monolithic LLM job; recent examples include RLHF/RLAIF training and agentic workflows. To cope with this shift, we propose FlowMesh, a multi-tenant service fabric that executes and optimizes these workloads as one shared service instead of isolated pipelines. It decomposes workflows into fine-grained operators with recorded lineage, enabling de-duplication of work across users and batching requests on the same hardware while preserving per-workflow provenance. A global control plane maintains a cluster-wide pool of ready operators and uses a single utility function to pick both the batch and the worker, balancing throughput, cost, and data locality on heterogeneous GPUs. The data plane is an elastic fleet of stateless workers backed by a content-addressable store, enabling rapid,  automatic scale-out, safe retry after preemption, and portability across managed clusters such as Kubernetes and geo-distributed GPU marketplaces such as Vast.ai.  Compared with baseline solutions, FlowMesh achieves up to 3.8× cost reduction and 2.0× lower energy usage, provides a similar or better latency profile, and remains efficient under dynamic and failure-prone conditions.
\end{abstract}
]


\printAffiliationsAndNotice{}  

\section{Introduction}
\begin{figure*}[t]
    \centering
    \includegraphics[width=0.9\linewidth]{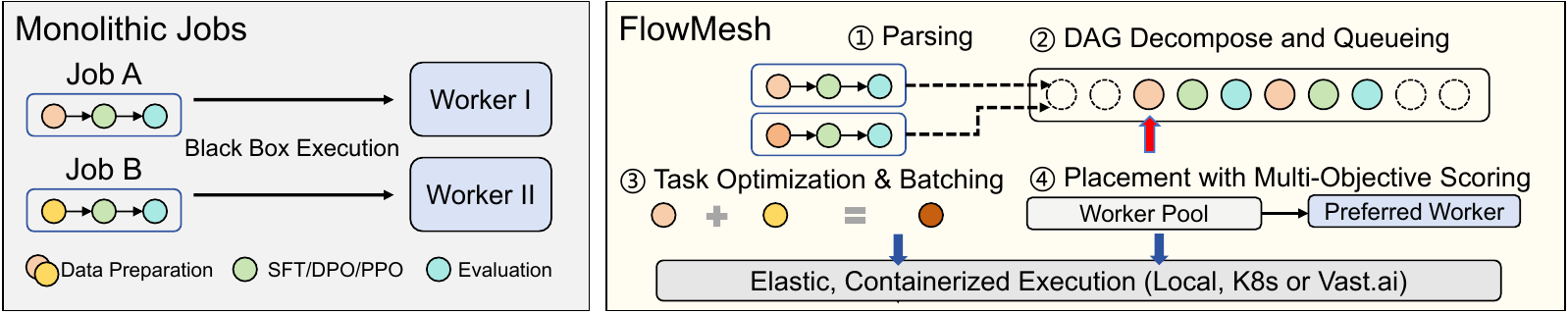} 
    \caption{
        FlowMesh decomposes multi-stage LLM workflows into fine-grained,
        tasks and dispatch them among distributed workers.}
    \label{fig:intro}
\end{figure*}

The dominant paradigm in applied machine learning is undergoing a significant transformation, shifting from monolithic, single-shot model training and inference to the orchestration of complex, multi-stage post-training workflows. These workflows, encompassing methodologies like Reinforcement Learning from Human Feedback (RLHF)~\cite{ouyang2022training}, LLM-as-a-Judge~\cite{gu2024survey}, Reinforcement Learning from AI Feedback (RLAIF)~\cite{lee2024rlaif}, multi-agent systems and agentic workflows~\cite{yao2023react}, are now the primary drivers of LLM research and production. Such processes are naturally expressed as Directed Acyclic Graphs (DAGs) of fine-grained, interdependent tasks, including data preparation, supervised fine-tuning, reward modeling, policy fine-tuning, and evaluation. This inherent graph structure facilitates decades of prior work on compilers and query optimizers for substantial system-level optimizations.

However, a critical mismatch exists between the graph-structured, multi-tenant nature of modern AI development and the execution models of contemporary cluster management and processing solutions. Platforms like Kubernetes and Spark, while proficient at managing containerized workloads, treat LLM jobs as opaque, black-box entities \cite{verma2015borg, zaharia2012spark}. They lack the semantic understanding of the internal DAG structure, precluding any form of cross-job optimization. This fundamental disconnect gives rise to three critical challenges that severely hinder efficiency, scalability, and cost-effectiveness in a GPU cluster or network: 

\emph{Redundant Compute:} In a typical multi-tenant system, multiple users submit jobs with similar workflows for concurrent execution, such as fine-tuning the base model on slightly varied datasets, or serving the same base model with different LoRA adapters \cite{hu2022lora}. Without a global awareness of the workload, existing systems compel each job to re-execute identical sub-tasks, like reward model inference or initial supervised fine-tuning (SFT), leading to a tremendous waste of expensive GPU cycles. 

\emph{Heterogeneous Infrastructures:} Production GPU clusters are inherently heterogeneous, comprising a mix of architectures (e.g., H100, A100) and varying interconnects. Naive scheduling fails to solve the complex, multi-objective optimization problem of matching diverse task requirements (e.g., CPU-bound data preprocessing, latency-sensitive inference, memory-bound training) to the most appropriate hardware. This mismatch also results in poor resource utilization, performance bottlenecks, and high operational costs. 

\emph{Non-trivial Provisioning:} The iterative, interactive nature of AI research and development generates highly bursty and unpredictable workloads. Statistically allocated resources, the norm in systems like Ray or Slurm-based setups \cite{moritz2018ray}, present a stark trade-off: overprovisioning leads to costly idle capacity, while underprovisioning results in prohibitive queueing delays that stifle productivity.

In a nutshell, the core issue is one of abstraction. Forcing a graph-structured workflow into a large, block-resource abstraction is inherently inefficient. It either necessitates over-provisioning for the peak resource needs of the entire workflow or introduces significant external orchestration complexity. An ideal system should consider inter-workflow dependencies and resource needs at a fine-grained, operator level, enabling optimizations that are impossible when the entire workflow is implemented in an opaque container.

In this paper, we propose FlowMesh, a service fabric that reframes LLM post-training workflows from isolated jobs into composable, multi-tenant services upon distributed GPU workers.  As demonstrated in Figure~\ref{fig:intro}, FlowMesh applies the following key ideas: (1) \emph{Workflow abstraction with containerized execution:} It encodes incoming workflows into DAGs, and uses a high-performance, containerized worker runtime to offset heterogeneity.  (2) \emph{Resource-aware placement and batching:} It uses a scheduler to make global placement and batching decisions that are topology- and resource-aware. (3) \emph{Serverless compute:} It dynamically and automatically adjusts the worker pool to match demand and offer fault tolerance.  

We implemented FlowMesh on a centralized Kubernetes GPU cluster as well as on a decentralized Vast.ai network~\cite{skywork2025vast}. Both backends run the same containerized workers pulled from a public registry and read/write data and logs to a shared content-addressable store. On Kubernetes, Pods are autoscaled and placed with GPU constraints; on Vast.ai, the control loop leases heterogeneous instances on demand and launches the identical image. Across both environments, the control plane schedules a queue of per-user DAGs, unifies identical operators, consolidates compatible ones into data locality-aware macro-batches, and delivers portable performance and cost efficiency without changing user workflows.
On realistic multi-tenant workloads, FlowMesh matches or exceeds baseline throughput while reducing costs by up to 3.8× and energy usage by up to 2.0× at similar or better latency, and it stays efficient under dynamic load and failures.

To conclude, contributions of this paper include: \vspace{-0.1in}
\begin{itemize}[leftmargin=*]
    \item We propose FlowMesh, a multi-tenant service fabric that runs LLM post-training and agentic workflows as a shared service instead of isolated pipelines.\vspace{-0.1in}
    \item FlowMesh breaks workflows into fine-grained operators with lineage and uses a global control plane plus an elastic pool of stateless workers to batch compatible work, schedule across heterogeneous GPUs, and scale across Kubernetes and decentralized GPU markets. \vspace{-0.1in}
    \item We evaluate FlowMesh on realistic multi-tenant workloads, and demonstrate significant performance improvements in throughput and cost efficiency.
\end{itemize}

\section{Background and motivation}
The motivation for FlowMesh stems from the massive economic cost of the status quo: high-end GPUs are a scarce and expensive resource, and a significant fraction of large cluster investments is wasted due to redundant computation and poor utilization. This inefficiency directly translates to wasted capital expenditure and reduced productivity. To ground our design, we first analyze the structure of modern ML workflows and then detail the shortcomings of existing systems to support these workflows.

\textbf{The Anatomy of LLM Post-Training Workflows}. To illustrate the inherent, composable graph structure and compute reuse opportunities, we deconstruct two canonical examples. 

\begin{table*}[t]
\begin{scriptsize}
\begin{center}
\caption{Comparison of off-the-shelf system solutions for LLM workflows.}
\label{tab:system-limitations}
\begin{tabular}{@{}lccccc@{}}
\toprule
\textbf{System Class} & \textbf{Examples} & \textbf{Cross-Workflow Reuse} & \textbf{Heterogeneity/Topology} & \textbf{Elastic Scaling} & \textbf{DAG} \\
 & & & \textbf{Awareness} &  & \textbf{Abstraction} \\
\midrule
Workflow Engines & Argo, Airflow & No (Per-job isolation) & Limited (Node affinity) & Limited & Yes (Defines jobs) \\
Data Systems & Ray, Dask & No (Single-job scope) & Partial (Actor placement) & No (Static cluster) & Yes (Dynamic tasks) \\
LLM Serving Systems & vLLM, SGLang & No (Per-request opt.) & No (Single-node focus) & Partial (Model-level) & No (Stateless endpoint) \\
Container Ochestrator & Kubernetes & No (Opaque pods) & Limited (No topology) & Yes (Pod-level) & No (Pod is primitive) \\
\midrule
\textbf{FlowMesh (Our Work)} & - & \textbf{Yes} & \textbf{Yes} & \textbf{Yes (Service-level)} & \textbf{Yes (Workflow graph)} \\
\bottomrule
\end{tabular}
\end{center}
\end{scriptsize}
\end{table*}

\begin{figure}[h]
    \centering

    \begin{minipage}{0.9\linewidth}
        \centering
        \includegraphics[width=\linewidth]{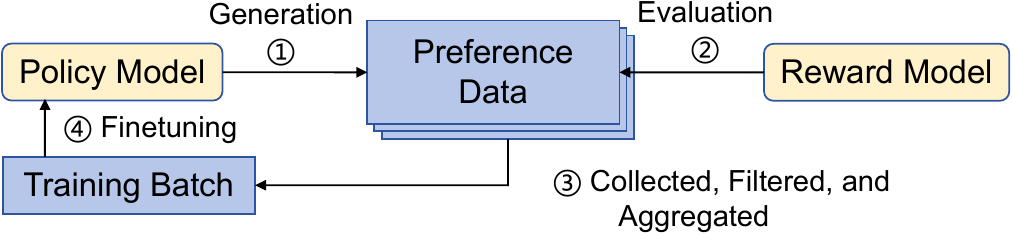}
    \end{minipage}

    \caption{RLHF/RLAIF through reward and feedback.}
    \label{fig:rlhf}
\end{figure}

\emph{RLHF Training:} A typical RLHF pipeline (shown in Figure~\ref{fig:rlhf}) is an iterative DAG consisting of several distinct stages. It begins with (1) parallel generation of responses from a policy model (often an initial SFT model), followed by (2) the scoring of these responses by a separate reward model. This preference data is then (3) collected, filtered, and aggregated into a training batch. Finally, (4) a distributed fine-tuning step, commonly using an algorithm like Proximal Policy Optimization (PPO) \cite{schulman2017ppo}, updates the policy model's weights. In a multi-tenant environment where multiple teams are iterating on variants of the same base model (e.g., Llama-3-8B), the SFT and reward model inference steps (1 and 2) are frequently identical across numerous concurrent experiments. These compute-intensive stages represent a prime target for deduplication.

\begin{figure}[h]
    \centering
    \begin{minipage}{0.9\linewidth}
        \centering
        \includegraphics[width=\linewidth]{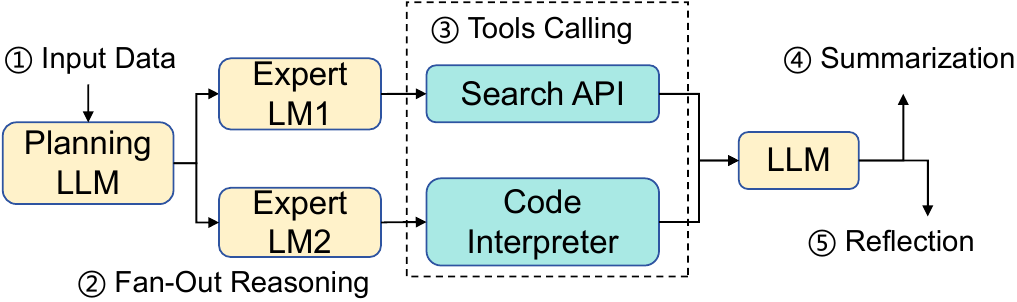}
    \end{minipage}

    \caption{An agentic workflow orchestrates reasoning, tool use, and reflection.}
    \label{fig:agenticworkflow}
\end{figure}

\emph{Agentic Workflows:} More complex agentic systems, such as those based on the ReAct paradigm \cite{yao2023react}, chain multiple model invocations, tool use, and reasoning steps (shown in Figure~\ref{fig:agenticworkflow}). For instance, an agent might involve a DAG where a planning LLM's output is parsed, used to call an external tool (e.g., a search API or a code interpreter), and the tool's output is then fed back into the LLM for summarization or reflection. In this architecture, the tool-use and summarization modules can be implemented as shared, stateless services shared by different agentic workflows, each pursuing distinct goals.

\textbf{Prior Solutions and Limitations}. To deploy these workloads and provide low-latency, low-cost service, prior solutions fail to address the specific, composite challenges of these new workloads. We categorize the shortcomings of four classes of existing systems in Table~\ref{tab:system-limitations}. 
Our analysis reveals that no existing solution holistically addresses the challenges. Workflow orchestrators define DAGs but lack multi-tenancy. Distributed frameworks like Ray are powerful for single, large jobs but operate within a static cluster allocation and do not share work across jobs. Specialized inference servers like vLLM and SGLang achieve high throughput for single models but are stateless endpoints unaware of the broader workflow context \cite{kwon2023efficient, zheng2023sglang}. Finally, the default Kubernetes orchestrator lacks the semantic understanding of LLM workflows to perform intelligent placement, especially concerning hardware heterogeneity and network topology.

\textbf{Motivation and Ideas}. Our analysis above distills into three foundational design principles for FlowMesh.

\emph{DAG abstraction with containerized execution.} Each tenant's workflow is parsed into a DAG, with each operator representing an LLM or algorithmic stage. Each operator is given a deterministic identity derived from its code, model/version, hyperparameters, and input lineage \cite{shen2025batchqueryprocessingoptimization}. This enables common-subgraph reuse: semantically identical sub-tasks (e.g., SFT on the same base checkpoint, reward-model inference over overlapping batches) are executed once and their materialized results are shared based on lineage, eliminating redundant computation and exposing reuse opportunities that job-scoped systems cannot see. To offset heterogeneity, we use a high-performance, containerized worker runtime; each local runtime wraps a state-of-the-art LLM runtime, such the vLLM engine that performs continuous batching. 

\emph{Resource- and topology-aware job scheduling} is performed at the operator granularity using a multi-objective cost model that accounts for compute demand, memory footprint, communication patterns, and SLOs. FlowMesh co-locates communication-heavy edges on high-bandwidth interconnects and binds model training to appropriate GPU classes. Crucially, it exploits \emph{semantic} similarity to batch across tenants: compatible LLM invocations (policy rollouts, reward scoring, evaluation passes) are continuously batched through the worker runtime, improving throughput and tail latency. Topology awareness (intra-node vs.\ cross-WAN) and heterogeneity handling (L40/A100 mix) are explicit in the placement logic.

\emph{Serverless compute for elasticity and resilience.} FlowMesh decouples the logical DAGs from physical resource capacity and manages worker pools elastically, scaling based on queue depths, cost targets, and SLOs. Tasks are lineage-backed, enabling deduplication, speculative re-execution, and fast recovery on preemption or failure; this borrows ideas from serverless and cloud-native computing to leverage stateless workers. Overall, FlowMesh's serverless control plane converts static, ill-provisioned deployments into an adaptive fabric that matches real-time demand to distributed workers underneath.

\section{System Design}
\begin{figure}
    \centering
    \includegraphics[width=0.9\linewidth]{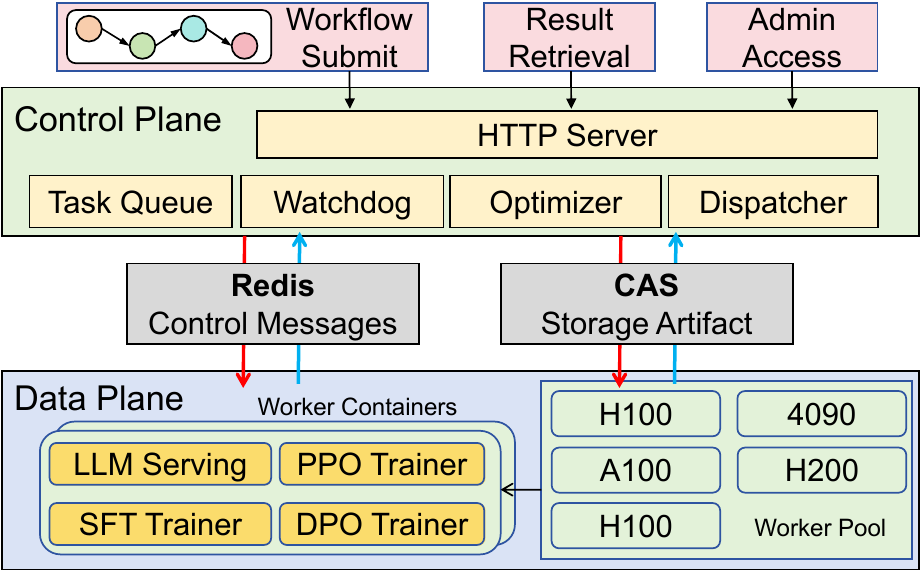}\vspace{-0.1in}
    \caption{\vspace{-0.1in}System overview of FlowMesh.}
    \label{fig:overview}
\end{figure}

\textbf{Overview}. As shown in Figure~\ref{fig:overview}, FlowMesh is architected as a dual-plane service fabric designed to improve the efficiency of multi-tenant, graph-structured LLM workflows with distributed workers. FlowMesh uses a DAG representation for the submitted workflow. Each operator in the workflow is assigned a unique, deterministic identity based on its model, parameters, and data lineage, enabling the system to automatically find and merge identical operators across different user workflows. These DAGs are managed by a centralized Control Plane, which employs a multi-objective, topology-aware scheduler that places operators on heterogeneous hardware and opportunistically batches compatible requests from different tenants to maximize GPU utilization. The execution itself is handled by a distributed Data Plane of stateless, containerized workers that are scaled elastically based on real-time demand. The combination of stateless workers and a lineage-backed DAG also provides inherent fault tolerance, allowing for simple, rapid recovery from failures. In the following sections, we elaborate on the details of each module. 

\textbf{DAG Abstraction}.
Prior work often models agentic workflows as DAGs, with each operator depicting LLM-based agent interaction, generic data processing or tool usage. Beyond this, we generalize this strategy to a broad class of LLM workflows, including operators for evaluation, data preparation, and RL-style post-training (e.g., SFT/DPO/PPO) \cite{rafailov2023dpo}. The system also maintains a {queue of per-request DAGs} for first-come-first-serve processing of the LLM workflows; each of them is compiled into its own DAG. Execution therefore proceeds by repeatedly taking ready operators from the DAGs in the queue and scheduling them when possible. By consolidating {similar} operators at dispatch time, the system achieves high utilization and low latency without compromising workflow isolation or provenance fidelity. This abstraction gives a clear lineage, reproducibility, and deduplication.


\textbf{Cross-DAG consolidation.} FlowMesh's scheduler looks across the frontiers of many queued DAGs and performs the following actions: 

\emph{Unification by identity (deduplication).} Each DAG is decomposed into fine-grained \emph{operators}, which are the unit of reasoning and reuse. Every operator then receives a deterministic identity over its full execution context $H_{\mathrm{task}}
= \texttt{hash}\!\big(
  H_{\mathrm{model}}, 
  \mathrm{canonical}(P), 
  H_{\mathrm{in},1,\cdots,n}
\big)$, where $H_{\mathrm{model}}$ digests the executor model, $\mathrm{canonical}(P)$ is a serialization of hyperparameters and resource hints, and $H_{\mathrm{in},i}$ hash the upstream datasets and prompts. If multiple ready operators from different DAGs share the same $H_{\mathrm{task}}$, the system executes the computation at most once (or skips it entirely if the output already exists) and {fans out} the resulting artifact to each consumer DAG. This eliminates redundant compute while preserving each workflow’s structure. 

\emph{Consolidation by execution signature (batching).} When operators are not byte-identical but are {compatible} for joint execution (e.g., same executor and parameters, different inputs that can be microbatched), they are scheduled together as one batched run. Compatibility is captured by an execution signature $H_{\mathrm{exec}} = \texttt{hash}\!\big(H_{\mathrm{model}}, \mathrm{canonical}(P), \mathrm{resource\_class}\big)$, which deliberately omits the input hashes. The batched run produces per-input results that are routed back to the corresponding edges in their original DAGs. This improves throughput without conflating provenance. In both cases, each DAG remains a first-class, isolated object; we only unify executions, not the workflow graphs themselves.

\textbf{Provenance and Isolation.}
All consolidations are gated by the above functions: \emph{exact-match} ($H_{\mathrm{task}}$) enables reuse while \emph{compatible-match} ($H_{\mathrm{exec}}$) enables batching inference tasks with the same LLM. Lineage is recorded per DAG edge, so prompts and intermediate outputs are referenced across workflows. Multi-tenant isolation also follows naturally from labeling the prompts and intermediate outputs and resolving the DAGs.

\subsection{Control Plane:  Routing and Online Batching}
\label{sec:control-plane}
The control plane is the global coordinator of FlowMesh to decide where each operator should execute. Specifically, the control plane is responsible for:
(i) placement on heterogeneous accelerators,
(ii) cross-tenant batching of compatible work, and
(iii) continuous admission of new requests into already-running executors.
A key goal here is that these decisions are made under a {single, unified objective}, instead of via separate stages.
At any instant, FlowMesh groups ready operators that invoke the same executor/model under the same hyperparameters and resource class. We denote this by $H_{exec}$ as mentioned earlier, which omits the specific input payloads. All operators that share a given $H_{\mathrm{exec}}$ are {batch-compatible}: they can run on the same long-lived executor with the same loaded weights. We denote the current compatible set by $S(H_{\mathrm{exec}})$.

\textbf{Scheduling Objective.}
For each worker $j$ and each candidate admitted batch $B \subseteq S(H_{\mathrm{exec}})$ that $j$ could serve, the control plane estimates a scalar utility 
\begin{equation} U(j, B) = w_t \, T_{\mathrm{eff}}(j, B) \;-\; w_c \, C(j) \;+\; w_\ell \, G_{\mathrm{loc}}(j, B).\label{eq:utility}\end{equation}

Each term reflects one dimension of system behavior:
\begin{itemize}
\item $T_{\mathrm{eff}}(j, B)$ is the predicted {effective throughput} of serving $B$ on worker $j$ (e.g., tokens/sec or samples/sec), given the model architecture, sequence lengths in $B$, and the GPU class of $j$. This pushes toward high accelerator utilization.
\item $C(j)$ is the normalized cost rate of keeping worker $j$ active (e.g., on-demand H100 vs.\ consumer GPU). This favors cheaper capacity when latency allows.
\item $G_{\mathrm{loc}}(j, B)$ is the {locality gain} from routing $B$ to $j$: it is higher if $j$ already has the needed model weights in GPU memory, the relevant adapters and tokenizers on local disk, or hot KV/cache state for that $H_{\mathrm{exec}}$. This rewards reuse and minimizes network traffic.
\end{itemize}

The weights $(w_t, w_c, w_\ell)$ encode cluster policy. Latency-sensitive interactive evaluation can downweight $C(j)$, preferring ``fast-but-expensive'' GPUs. Large offline rollout generation can do the opposite and favor cheaper hardware, even at lower tokens/sec, as long as deadlines are met.
Note that a pair $(j,B)$ is \emph{feasible} only if it satisfies hard constraints:
(1) VRAM / memory bandwidth limits on $j$, (2) architectural requirements (e.g., instruction set, GPU generation), and (3) tenant affinity/anti-affinity (e.g., ``must run on private GPUs'').
Among feasible candidates, the control plane prefers the $(j,B)$ that maximizes $U(j,B)$ in Eq.~\ref{eq:utility}.
Lastly, since the worker's admission queue $Q_j(H_{\mathrm{exec}})$ is shared, requests from different tenants that invoke the same model and parameters naturally co-reside on the same worker. 
This enables cross-DAG consolidation mentioned earlier in this section.

\textbf{Online Admission and Continuous Batching.}
FlowMesh does not treat $B$ as a one-shot job batch. Instead, each worker $j$ maintains a live admission queue $Q_j(H_{\mathrm{exec}})$ for every execution signature it is currently serving. 
The control plane continuously feeds new compatible operators from $S(H_{\mathrm{exec}})$ into $Q_j(H_{\mathrm{exec}})$, subject to feasibility and utility. In other words, $B$ is the next slice of work to admit into an already running executor on $j$. This turns scheduling into continuous serving: multiple tenants' requests stream into persistent executors that already have the correct model weights resident, and the runtime on $j$ performs dynamic (token-by-token / step-by-step) batching across those requests. GPU occupancy stays high, and latency stays low, because the control plane can admit work immediately into a hot lane rather than waiting to form a large static batch or spin up a fresh worker.

When the control plane admits work to $Q_j(H_{\mathrm{exec}})$, it sends the operator spec(s), input content hashes, and the expected output schema.
The worker's continuous-batching runtime executes these requests, produces one output per logical input, and writes each output to the shared content-addressable store (CAS) under a fresh content hash. The control plane records those hashes and unblocks the downstream edges in each originating workflow DAG. Because every intermediate result is content-addressed, provenance is explicit.

\textbf{Remark}. 
Adding up the above designs, the control plane turns all tenant workflows into a single global stream of ready operators, continuously routes compatible requests into persistent executors, and uses one utility function (Eq.~\ref{eq:utility}) to trade off throughput, cost, and data locality. It is simultaneously a scheduler, a batcher, and a multi-tenant serving router.

\subsection{Data Plane: Elastic, Stateless Execution}
\label{sec:data-plane}
The data plane is an elastic pool of stateless worker containers plus a shared content-addressable store (CAS) \cite{benet2014ipfscontentaddressed}. Each worker behaves like a {persistent serving executor} for one or more execution signatures $H_{\mathrm{exec}}$: once it has loaded the relevant model weights (and, if applicable, adapters, tokenizers, KV caches, etc.), the control plane keeps streaming compatible requests to it. Workers expose a continuous-batching runtime that can interleave requests from different tenants in real time.

\textbf{Auto-scaling.}
Because workers are logically stateless, FlowMesh can scale them directly with demand: (1) If global queue depth or SLO pressure rises for some $H_{\mathrm{exec}}$, new workers that can serve that signature are provisioned (e.g., new Pods in Kubernetes or newly leased GPU instances). (2) If demand falls and workers go idle, surplus workers are retired to avoid paying for idle accelerators.
Joining the pool means starting to act as a serving lane for $H_{\mathrm{exec}}$: the worker loads the needed model weights, initializes its continuous-batching runtime, and begins accepting admissions from the control plane. Leaving the pool means draining outstanding requests, flushing outputs to CAS, and exiting.
If a worker fails mid-flight (e.g., preemption), any in-progress operators that did not complete are simply marked ready again and can be re-admitted elsewhere; completed outputs remain in CAS to simplify preemption.

\textbf{CAS and Data Locality.}
All data artifacts, i.e., model checkpoints, adapters, tokenizers, rollout samples, reward scores, evaluation traces, summaries, live in an S3-compatible content-addressable store. Each artifact is named by a hash of its bytes. A worker serving $H_{\mathrm{exec}}$ operates as follows: (1) The control plane hands it new requests by reference, i.e., via content hashes of required inputs. (2) The worker then resolves any missing inputs: it first checks its local cache (for weights, adapters, and recent inputs), then lazily fetches uncached artifacts from CAS.
(3) The worker's continuous-batching runtime executes those requests, potentially interleaving many tenants' operators in a single forward/backward pass.
(4) For each logical input, the worker emits an output artifact and writes it back to CAS under a new content hash. 

Our solution has the following goals. 
First, locality emerges naturally: once a worker is ``hot'' for some $H_{\mathrm{exec}}$, it tends to keep the right weights, KV cache state, and tokenizers resident, so future compatible requests are both faster (no reload) and cheaper (no repeated multi-GB transfers). The control plane captures this via $G_{\mathrm{loc}}(j,B)$ in Eq.~\ref{eq:utility} and preferentially routes more of that $H_{\mathrm{exec}}$ traffic to that worker.
Second, provenance is automatic: every intermediate is content-addressed, so downstream workflow stages receive immutable hashes rather than opaque in-memory pointers. This preserves isolation between tenants while still allowing cross-tenant reuse of identical results.

\textbf{Remark}. 
To conclude, the data plane is an autoscaled pool of persistent executors, each one a continuous-batching serving lane for one or more $H_{\mathrm{exec}}$, backed by a shared CAS. Workers can appear, warm up, absorb multi-tenant traffic under tight SLOs, and disappear, without breaking lineage or forcing tenants to rewrite their workflows.

\begin{table*}[h!]
\begin{footnotesize}    
\centering
\caption{Mapping of FlowMesh Components to Infrastructure Primitives}
\label{tab:implementation}
\begin{tabular}{@{}lll@{}}
\toprule
\textbf{Architectural Component} & \textbf{Implementation on A Centralized K8s Cluster} & \textbf{Implementation on Decentralized Vast.ai} \\ \midrule
\textbf{Control Plane} & Kubernetes \texttt{Deployment} with custom controller & Runs on a stable, on-demand Vast.ai instance \\
\textbf{Worker Provisioning} & \texttt{HorizontalPodAutoscaler} (HPA) & Custom controller making API calls \\
\textbf{Task Dispatch} & Controller creates a Kubernetes \texttt{Pod} & Control plane sends task spec via SSH/startup script \\
\textbf{Storage (CAS)} & \texttt{PersistentVolume} backed by cluster storage & External S3-compatible object store \\
\textbf{Heterogeneity Info} & Kubernetes node labels & Real-time query results from Vast.ai \\
\textbf{Cost Model Input} & Pre-configured node costs & Dynamic, real-time prices from Vast.ai \\ \bottomrule
\end{tabular}
\end{footnotesize}
\end{table*}

\subsection{System Resilience}
\label{sec:resilience}

Resilience in FlowMesh emerges from three design choices: workers are stateless, and data is immutable and content-addressed. 
Instead of bespoke checkpointing, the system relies on idempotent re-execution with provenance that guarantees correctness. This follows the design principal of cloud-native computing. 
Each workflow remains its own DAG, while the control plane tracks per-DAG operator state (\texttt{READY}, \texttt{RUNNING}, \texttt{COMPLETED}) and exposes a pooled ready-operator queue across DAGs for scheduling and consolidation. When a worker fails or is preempted, any \texttt{RUNNING} operators it held are atomically returned to \texttt{READY}. A healthy worker then retrieves the same immutable inputs from the CAS and recomputes the result, making retry the common recovery path.
Before re-executing a retried operator, the control plane checks whether the operator’s output already exists in the CAS, possibly produced by a consolidated run that served multiple DAGs. If it does, execution is skipped and the cached artifact is fanned back to the operator’s original edges, preserving provenance while eliminating duplicate work. 
Because workers carry no durable state and all inputs/outputs are CAS-backed, FlowMesh comfortably leverages preemptible or transient capacity. Autoscaling policies absorb bursts and drain lull periods without correctness risk, turning inexpensive but unreliable instances into an operational advantage.

To mitigate stragglers and long-tail variance, the scheduler may launch speculative replicas. Only the first successful completion is published to the CAS; late or duplicate completions are discarded by content identity. This delivers at-most-once publication even under at-least-once execution.
All fan-in and fan-out are expressed via content hashes, and each per-DAG edge records the exact versions consumed and produced. The result is precise prospective and retrospective provenance, enabling audit and exact replay.

\section{Implementation}

To demonstrate the flexibility of its core abstractions, FlowMesh is implemented to run on two fundamentally different infrastructure paradigms: a centrally managed Kubernetes cluster, typical of enterprise environments, and a decentralized, peer-to-peer compute marketplace, represented by Vast.ai. Table~\ref{tab:implementation} summarizes the mapping of FlowMesh components to the primitives of each environment.

\vspace{0.05in}\noindent\textbf{Deployment on Kubernetes Clusters}. In a Kubernetes environment, FlowMesh is implemented as a set of custom controllers and a scheduler plugin, integrating natively with the cluster's control plane \cite{portworx2024k8sai}.
The Planner, Router, and scheduler logic are packaged into a single Go-based binary that runs as a Kubernetes \texttt{Deployment}. It watches for \texttt{FlowMeshWorkflow} Custom Resource Definitions (CRDs) submitted by users. The scheduler is implemented as a Kubernetes scheduler plugin. When the controller identifies a ready task, it creates a placeholder \texttt{Pod}. The FlowMesh plugin intercepts this \texttt{Pod} and executes its Filter/Score logic to determine the optimal placement. The worker runtime is packaged as a Docker container and managed by a Kubernetes \texttt{Deployment}. A \texttt{HorizontalPodAutoscaler} is configured to scale the number of worker \texttt{Pods} based on the length of the global ready-task queue, exposed as a custom metric via Prometheus. The CAS is backed by a \texttt{PersistentVolumeClaim} that can be satisfied by any cluster-wide storage solution, such as CephFS, or a cloud provider's object store.

\vspace{0.05in}\noindent\textbf{Deployment on the Vast.ai Network}. Deploying on Vast.ai requires adapting FlowMesh to a dynamic, heterogeneous, and geographically distributed P2P marketplace \cite{skywork2025vast}.
The control plane runs on a single, stable, on-demand Vast.ai instance. It uses the official Vast.ai Python CLI and SDK to interact with the marketplace programmatically \cite{vastai2024cli, vastai2024sdk}. The Kubernetes HPA is replaced by a custom control loop. This loop periodically queries the ready-task queue. To scale up, it formulates a query based on a task's requirements and executes \texttt{vastai search offers} to find suitable instances. The scheduler's cost model directly ingests the dynamic pricing and host reliability metrics returned by the API. It then rents the best-scoring machine using \texttt{vastai create instance}. To scale down, the loop identifies idle workers and terminates their instances using \texttt{vastai destroy instance}. Since Vast.ai instances have ephemeral storage, the CAS must be an external, S3-compatible object store. When dispatching a task to a newly created worker, the control plane passes the task specification, including input artifact hashes and temporary credentials for the object store, via environment variables or a startup script configured in the instance creation command \cite{vastai2024vms}.

\vspace{0.05in}\noindent\textbf{Containerized Workers}. Each worker is a Docker image that wraps a state-of-the-art engine (e.g., vLLM) behind a thin agent: it pulls inputs by content hash from the CAS, executes, and pushes immutable outputs, keeping workers stateless and interchangeable. On Kubernetes, the same image runs as a deployment and scales with queue depth; placement filters on GPU type/VRAM. On Vast.ai, the control loop leases instances, pulls the identical image, and starts the container with read-only CAS credentials. A single CUDA~12.8 image supports heterogeneous GPUs: GPU-aware runtimes (PyTorch) select kernels by compute capability; devices are injected via \texttt{nvidia-container-runtime}, covering A100/H100 and prosumer RTX GPUs. Images are published to a private registry with regional mirrors; Kubernetes pulls via \texttt{ImagePullSecrets} and Vast.ai via short-lived tokens, while the S3-compatible CAS remains the source of truth for weights, adapters, tokenizers, and intermediates.


\section{Evaluation}
We conduct an empirical evaluation to justify FlowMesh's design and evaluate its performance. Our experiments are guided by the following questions: (1) \label{enum:rq1} How does FlowMesh perform in serving concurrent LLM workflows compared to prior solutions? (2) \label{enum:rq3} How does each component contribute to FlowMesh, and how well does it behave in various deployment scenarios, including failures? (3) \label{enum:rq5} How does FlowMesh scale and integrate with different runtime backends? 

In this section, we first analyze throughput using primitive workflows, followed by cost and energy usage comparisons. We then present ablation and robustness analyses, and conclude with scalability experiments.

\subsection{Experimental Setup}
\vspace{0.05in}\noindent\textbf{LLM Workflows}. We use four primitive LLM post-training workflows spanning common deployment tasks.\vspace{-0.05in}

\begin{itemize}[leftmargin=0.2in, itemsep=1pt, topsep=1pt]
    \item \emph{Agentic (Inference) Workflow:} multiple agentic workflows covering various model and dataset configurations.\vspace{-0.05in}
    \item \emph{Supervised Fine-Tuning (SFT) Workflow} for both LoRA SFT and full-weights SFT, followed by an evaluation pipeline and a hold-out test set.\vspace{-0.05in}
    \item \emph{Direct Preference Optimization (DPO) Workflow} for RL training using configured preference pairs, followed by an evaluation job. \vspace{-0.05in}
    \item \emph{Proximal Policy Optimization (PPO) Workflow} for RL training, followed by an evaluation job.
\end{itemize}

We use public datasets including GSM8K \cite{cobbe2021gsm8k}, MMLU \cite{hendryckstest2021}, and TruthfulQA \cite{lin2022truthfulqameasuringmodelsmimic} in these workflows.

\textbf{Test Workloads} are divided into two groups:\vspace{-0.05in}

\begin{itemize}[leftmargin=0.2in, itemsep=1pt, topsep=1pt]
\item\emph{Group A} consists of five agentic inference workflows with diverse DAG topologies, including reasoning chains, retrieval-augmented generation, and multi-agent collaboration tasks. These workflows are designed to evaluate FlowMesh’s ability to serve heterogeneous inference pipelines efficiently. We sample 200 workflows from these datasets to construct the test workload.\vspace{-0.05in}

\item\emph{Group B} extends Group A by adding four post-training pipelines following RLHF/RLAIF methodologies. Each pipeline comprises typical sub-stages such as SFT, PPO, and DPO, enabling us to assess scheduling, dependency management, and resource reuse under multi-stage fine-tuning workloads. We sample 200 workflows for training and 100 workflows for evaluating inference from GSM8K, respectively, to construct the test workload. 
\end{itemize}

\vspace{0.05in}\noindent\textbf{Baselines}. We compare FlowMesh against several baselines; they are categorized in whether to break down the workflow (decompose or monolithic) and how to schedule the jobs (first-fit, static routing, or consolidate):\vspace{-0.05in}

\begin{itemize}[leftmargin=0.2in, itemsep=1pt, topsep=1pt]
    \item \emph{MF: Monolithic + First-Fit:} We use vLLM + TRL~\cite{Kwon2023vLLM, vonWerra2023TRL} and implement a simple first-fit heuristic to place the next entire workflow on the first available worker. \vspace{-0.05in}

    \item \emph{DS: Decompose + Static:} We implement  JellyBean~\cite{wu2022serving} that similarly decomposes complex ML workflows into operators and allocates them onto distributed workers using a static routing based on the operator's functionality; e.g., inference and SFT workers always do their designated jobs. \vspace{-0.05in}

    \item \emph{DR: Decompose + Round-Robin:} similar to DS, we decompose ML workflows but allocate the operators using a round-robin assignment to all available workers.\vspace{-0.05in}

    \item \emph{FlowMesh: Decompose + Consolidate:} Our solution decomposes the workflows, consolidates jobs by the operators, and allocates them to distributed workers.
\end{itemize}


The workflows utilize base models including Llama-3.1-8B, Llama-3.2-3B, and Llama-3.2-1B~
\cite{meta_llama3_2024}. The experiments are conducted on a local cluster (Section~\ref{exp1}, \ref{exp2}), Kubernates and Vast.ai (Section~\ref{exp3}). All the baselines and workers are implemented with vLLM for inference and Huggingface TRL for RL training~\cite{Kwon2023vLLM, vonWerra2023TRL}.  The batch size is set to 24 for Group A and 12 for Group B, respectively.

\subsection{Performance and Cost Analyses} 
\label{exp1}

Cloud-based serving platforms inherently balance {cost efficiency}, {monetary cost}, and {serving performance}.  
We first evaluate FlowMesh from the perspective of cost and energy efficiency, and then analyze its end-to-end serving latency and throughput scalability.

\paragraph{Experiment Metrics and Setup.}
Following \cite{rolof}, we assess both {monetary cost} and {energy consumption} under realistic workloads, and use two composite metrics, the {Cost–Delay Product (CDP)} and the {Energy–Delay Product (EDP)}, to jointly measure the trade-offs between efficiency and latency: $\text{CDP} = \frac{\text{Total Cost}}{\text{\#Tasks}} \times \text{AvgTime}, \text{EDP} = \frac{\text{Total Energy}}{\text{\#Tasks}} \times \text{AvgTime},$ where lower values indicate more cost- and energy-efficient results. We also report {average task latency} that aggregates all sub-stages within a workflow DAG (e.g., PPO fine-tuning, reasoning chains).

The experiment is conducted on 6 GPU workers, including Nvidia~H100~NVL~(94\,GB), RTX~4090~(48\,GB), and RTX~4090~(24\,GB) in equal proportion. Unit prices (\$/hour) reflect average rental rates from \href{https://vast.ai}{Vast.ai} in Oct 2025.  Each worker samples instantaneous power through periodic heartbeats, while the host integrates power over time to compute per-task energy usage.  To simulate realistic peak-and-tail traffic, we apply an exponentially decaying query rate from 6~to~0.6~qpm, with request workflows drawn randomly from Group~A workloads.

\begin{figure}[t]
    \centering
    \begin{minipage}{0.49\linewidth}
        \centering
        \includegraphics[width=\linewidth]{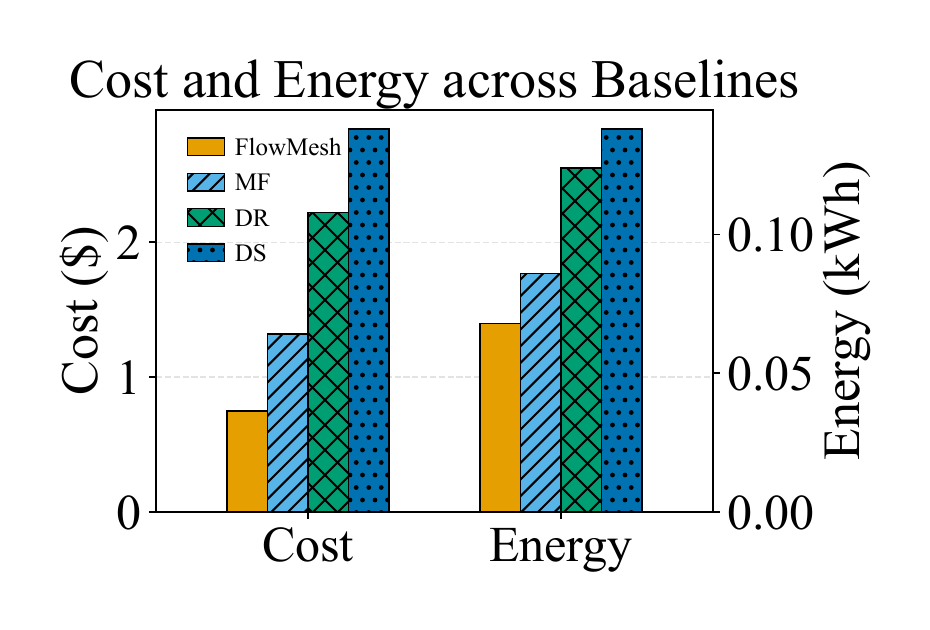}
        \vspace{-1mm}
    \end{minipage}
    \hfill
    \begin{minipage}{0.49\linewidth}
        \centering
        \includegraphics[width=\linewidth]{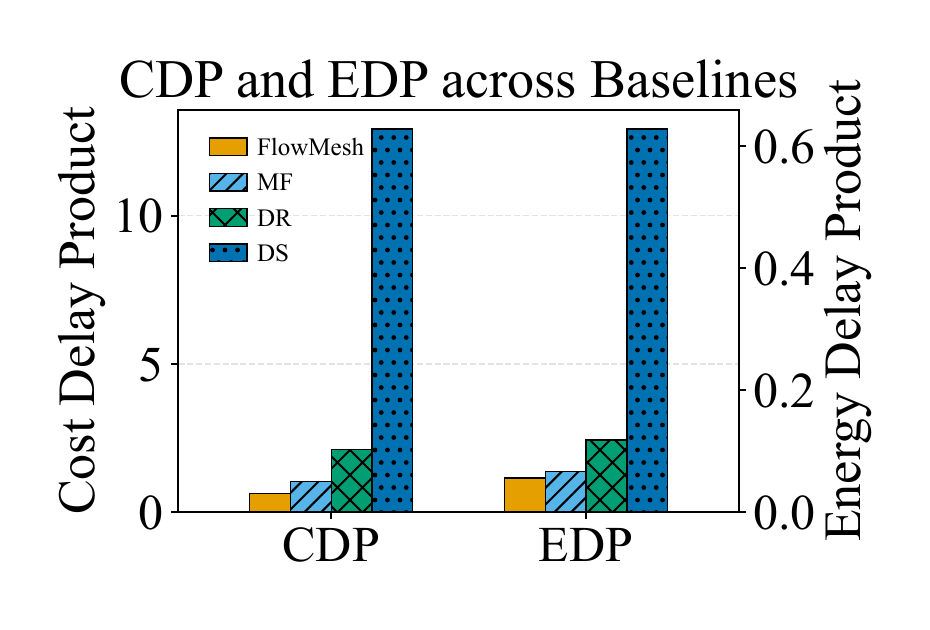}
        \vspace{-1mm}
    \end{minipage}
    \vspace{-5mm}
    \caption{
        {FlowMesh} compared with baselines.
        Left:~Total cost and energy consumption under identical workload conditions.
        Right:~Cost–Delay Product (CDP) and Energy–Delay Product (EDP).
    }
    \label{fig:flowmesh-cost-cdp}
\end{figure}

\begin{figure}[t]
    \centering
    \begin{minipage}{0.49\linewidth}
        \centering
        \includegraphics[width=\linewidth]{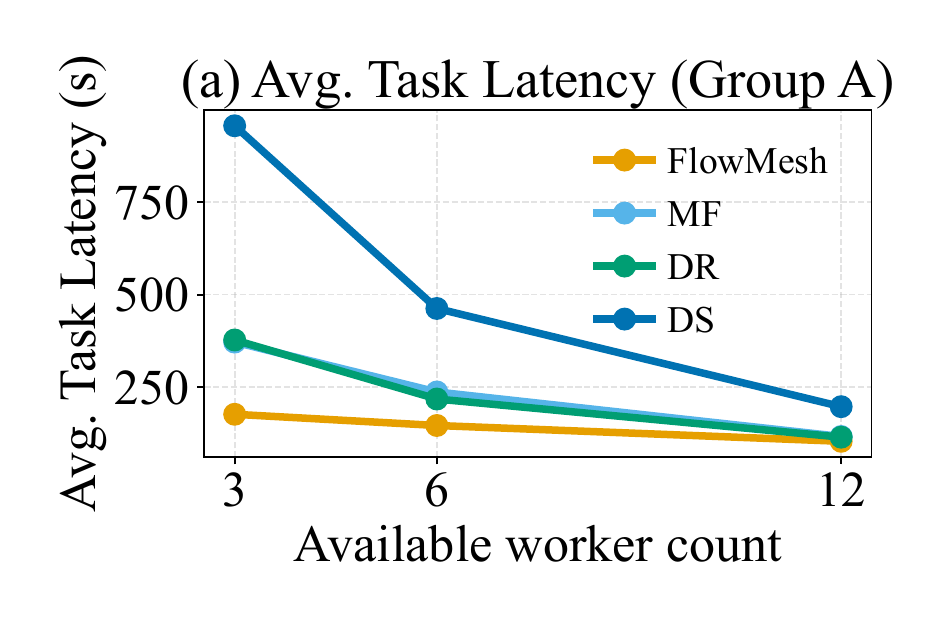}
        \vspace{-1mm}
    \end{minipage}
    \hfill
    \begin{minipage}{0.49\linewidth}
        \centering
        \includegraphics[width=\linewidth]{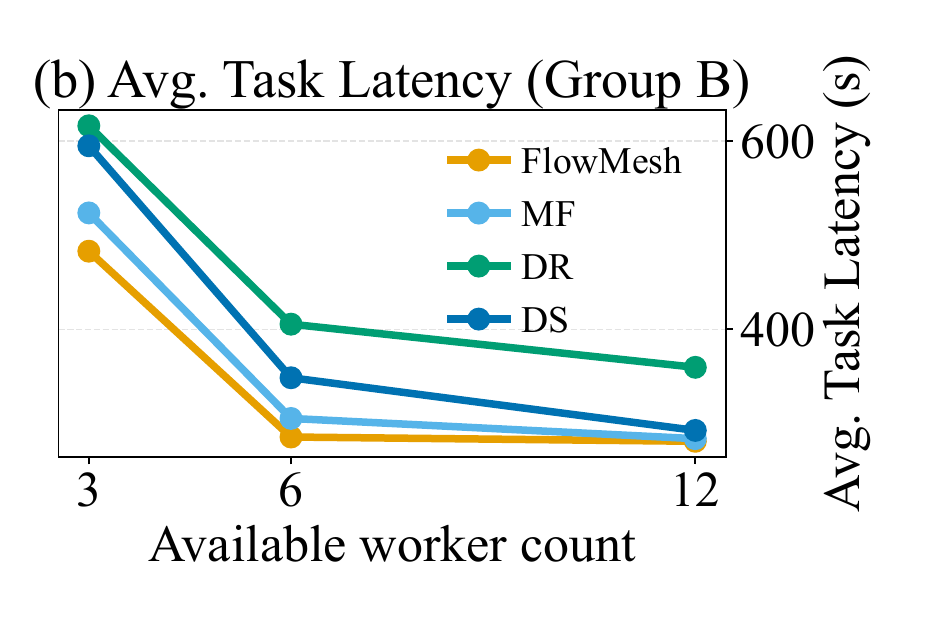}
        \vspace{-1mm}
    \end{minipage}
    \vspace{-5mm}
    \caption{
        Average task latency with different numbers of available workers. FlowMesh achieves similar or better latency in all cases. 
    }
    \label{fig:latency}
\end{figure}

\textbf{Comparisons.}
\autoref{fig:flowmesh-cost-cdp} shows that FlowMesh delivers the lowest cost and energy use among all baselines. It reduces monetary cost by 1.8×–3.8× and energy usage by 1.3×–2.0×, while achieving 2×–10× better CDP and EDP. Although DS attains slightly higher throughput by favoring H100 GPUs, FlowMesh instead makes better use of smaller GPUs, trading some throughput for substantially improved cost and energy efficiency. 

Results in Figure~\ref{fig:latency} indicate that FlowMesh consistently matches or outperforms all baseline schedulers in latency. The advantage is more significant with fewer workers, where baselines often queue for resources, while FlowMesh consolidates common tasks across workflows, thus skipping the queue. As more workers become available, everyone’s latency improves, the queueing pressure eases, and the gap narrows; this is as expected. Even then, FlowMesh maintains comparable or slightly better latency.


Overall, our experiments verify that FlowMesh achieves better cost and energy efficiency compared to baseline solutions; meanwhile, it keeps the latency comparable or even better relative to running the workflows as a whole piece. These gains stem from elastic resource allocation and improved scheduling.

\subsection{Ablation and Robustness  Study}
\label{exp2}
\textbf{Ablation Study.} As shown in~\autoref{tab:ablation_eff_cost}, we ablate three components in FlowMesh to quantify their contributions, thus justifying the design of our solution.

\emph{Operator Consolidation.}
Batching and executing homogeneous sub-tasks amortize repeated setup and warm-up (e.g., model/context loading).
On a batch of 24 concurrent agent workflows with the device mix in \S\ref{exp1}, 
{disabling} task consolidation increases average latency by \(1.36\times\) and total running cost by \(1.25\times\).
Benefits grow with higher concurrency as merge opportunities increase.

\emph{Elastic Resource Usage.}
Adapting the worker pool to the actual workload (pausing idle workers, reactivating on demand) cuts long-tail waste. Under the same setup, {disabling} elastic resource usage increases average latency by \(1.21\times\) and total cost by \(1.78\times\).

\emph{Multi-Objective Scheduling.}
A customizable objective (cost-first vs.\ performance-first) matches ready tasks to eligible workers. Compared with Round-Robin with random allocation, our full solution improves service efficiency by \(1.33\times\) and reduces cost by \(1.24\times\). We observe larger gains when the objective can be tuned specifically to each testing workload, although we did not report this here.

 
\textbf{Robustness.}
We verify the robustness of FlowMesh on heterogeneous, distributed hardware with two representative failure scenarios using the \S\ref{exp1} setup.

\begin{table}[t]
\begin{footnotesize}    
\begin{center}  
\caption{Ablation study results showing latency and cost improvements of each component.}
\begin{tabular}{lcc}
\toprule
\textbf{Disable Component} & \textbf{Latency ($\times$)} & \textbf{Cost ($\times$)} \\
\midrule
Operator Consolidation & \(1.36\times\) & \(1.25\times\) \\
Elastic Resource Usage & \(1.21\times\) & \(1.78\times\) \\
Multi-Objective Scheduling & \(1.33\times\) & \(1.24\times\) \\
\midrule
FlowMesh (full) & \(1.00\times\) & \(1.00\times\) \\
\bottomrule
\end{tabular}
\label{tab:ablation_eff_cost}
\end{center}  
\end{footnotesize}
\end{table}

\begin{table}[t]
    \centering
    \caption{Robustness test under two representative scenarios.}
    \label{tab:robustness_study}
    \begin{footnotesize}
        \begin{tabular}{lcc}
            \toprule
            \textbf{Error} & \textbf{Avg. Latency $\uparrow$} & \textbf{Detection Time} \\
            \midrule
            Worker Crash & 13.3\% & 30.0s \\
            Wrong Resource Spec & 5.1\% & 8.6s \\
            \bottomrule
        \end{tabular}
    \end{footnotesize}
\end{table}

\emph{Worker Crash.}
To emulate unexpected outages (e.g., host, network, or power faults), we forcibly take one H100 worker offline after 2 minutes. FlowMesh uses a \emph{watchdog} that couples worker health monitoring with dispatch. Each worker is expected to emit periodic heartbeats; when heartbeats are missed for an entire watchdog period (30s), the system marks the worker as failed, immediately re-enqueues any tasks that were in flight on that worker, and reschedules their dependent sub-tasks. The results show that FlowMesh continues the execution and completes all affected tasks with only a moderate increase in end-to-end latency (Table~\ref{tab:robustness_study}).

\emph{Wrong Resource Specification.}
Users specify resource demand (e.g., GPU memory requirement) using hint annotations to guide the placement.
We intentionally under-specify the GPU memory requirement in a multi-stage inference workflow. Similarly, the worker proactively reports the failure due to a resource shortage.
The control plane resubmits the faulty task to the next available worker with sufficient resources. As a result, the workflow still completes successfully, and only a small latency loss is observed.


\subsection{Scalability Study}
\label{exp3}

\begin{figure}[t]
    \centering
    \begin{minipage}{0.49\linewidth}
        \centering
        \includegraphics[width=\linewidth]{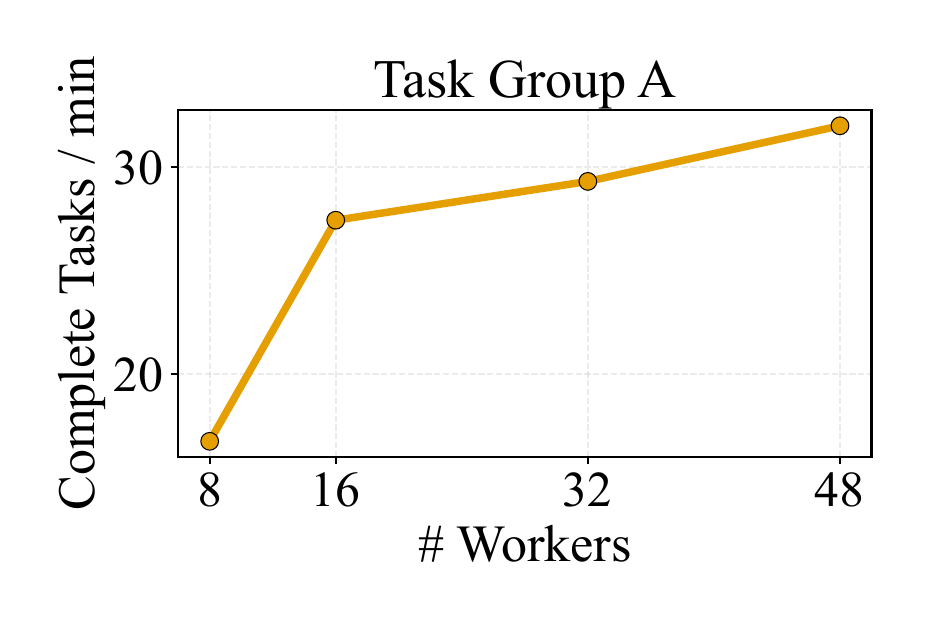}
        \vspace{-1mm}
    \end{minipage}
    \hfill
    \begin{minipage}{0.49\linewidth}
        \centering
        \includegraphics[width=\linewidth]{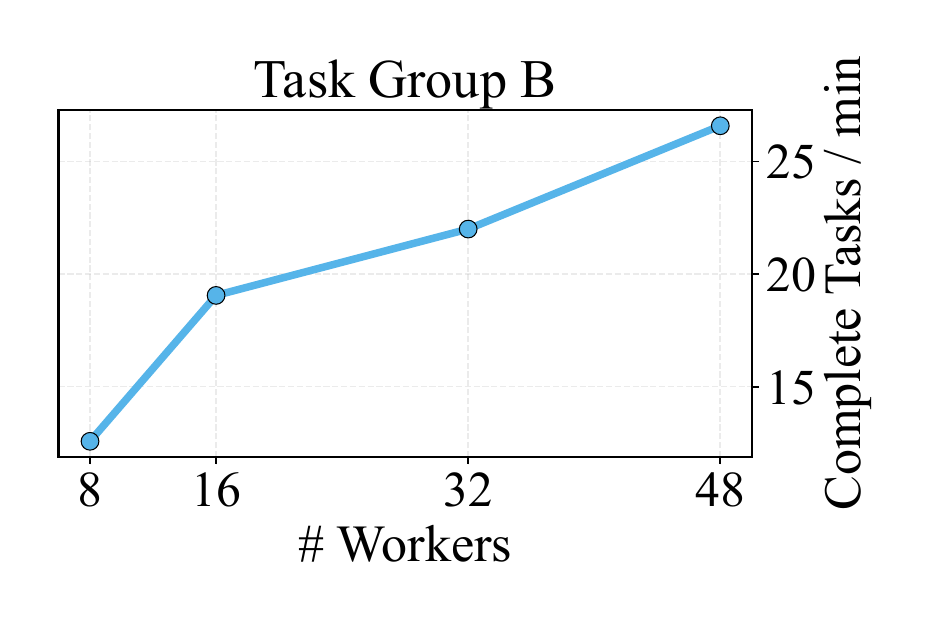}
        \vspace{-1mm}
    \end{minipage}
    \vspace{-5mm}
    \caption{
        Complete tasks under different numbers of workers on a Kubernetes cluster.
    }
    \label{fig:throughput}
\end{figure}
\begin{figure}[t]
    \centering
    \begin{minipage}{0.49\linewidth}
        \centering
        \includegraphics[width=\linewidth]{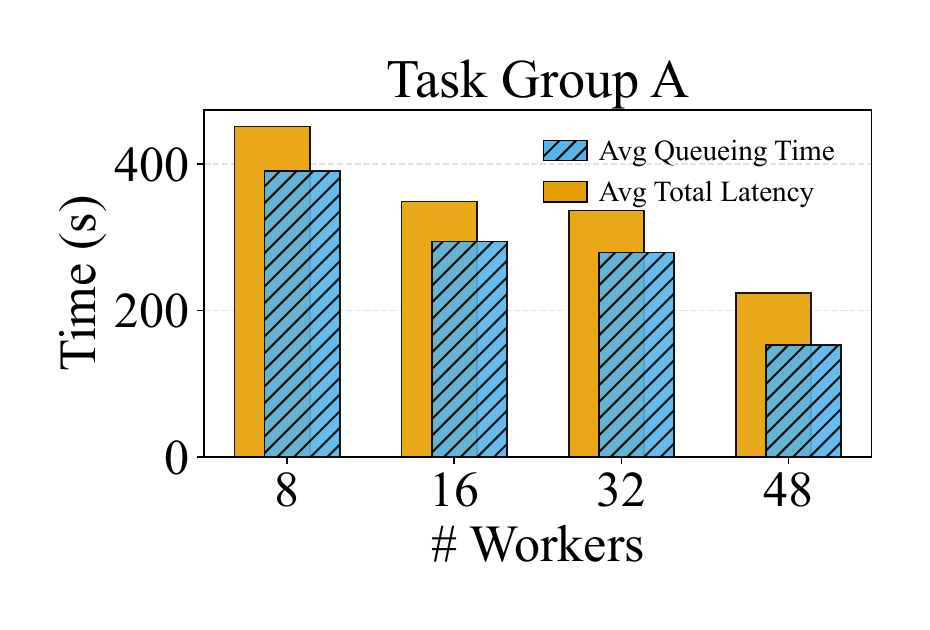}
        \vspace{-1mm}
    \end{minipage}
    \hfill
    \begin{minipage}{0.49\linewidth}
        \centering
        \includegraphics[width=\linewidth]{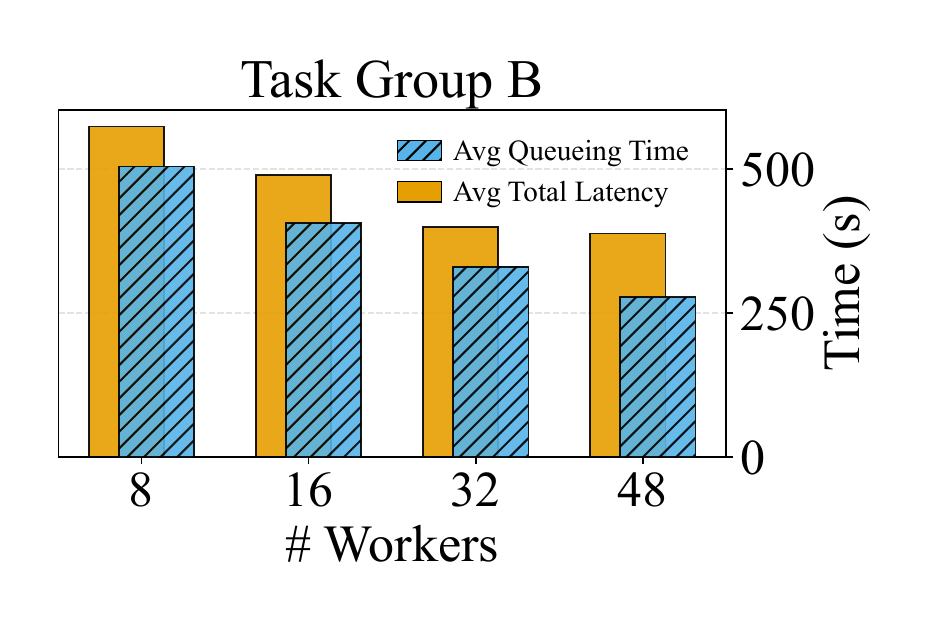}
        \vspace{-1mm}
    \end{minipage}
    \vspace{-3mm}
    \caption{
        Average queueing time and total latency under different numbers of  workers on a Kubernetes cluster.
    }
    \label{fig:scalability}
\end{figure}

\textbf{Scalability on Kubernetes.} To evaluate FlowMesh's scalability on large-scale clusters, we use the experiment in \S\ref{exp1} on a centralized Kubernetes cluster. The experimental cluster consists of up to 48~GPU nodes, each with one NVIDIA~H100 (80\,GB HBM3) GPU. The cluster is orchestrated using \texttt{kubectl}~v1.32.1 (Kustomize~v5.5.0). The workload remains the same, while the number of concurrent submissions is scaled to 64 to simulate a high-concurrency serving scenario with multiple tenants. 

Figure~\ref{fig:throughput} shows that increasing the number of workers improves the throughput sub-linearly. Specifically, in Group A, FlowMesh completes 17 tasks per minute from the queue with 8 workers, while completes 32 tasks per minute with 48 workers; in Group B, 13 tasks with 8 workers and 26 tasks per minute with 48 workers. On the other hand, Figure~\ref{fig:scalability} shows that increasing the number of workers reduces overall latency as the cluster scales. Specifically, FlowMesh reduces the queuing time, e.g., from 400 sec with 8 workers to 150 sec with 48 workers for Group A workload, while offering a similar per-workflow latency profile. 

In both scenarios, we observe a similar situation as in Figure~\ref{fig:latency}; queuing delay accounted for a large portion of the entire processing time. 
Indeed, the overhead caused by Kubernetes is still sizable, but improving Kubernetes and adapting FlowMesh to it falls outside the scope of this paper; we will defer this to future work. Nevertheless, these results indicate that FlowMesh can scale on a larger cluster. 

\begin{figure}[t]
    \centering
    \includegraphics[width=1.0\linewidth]{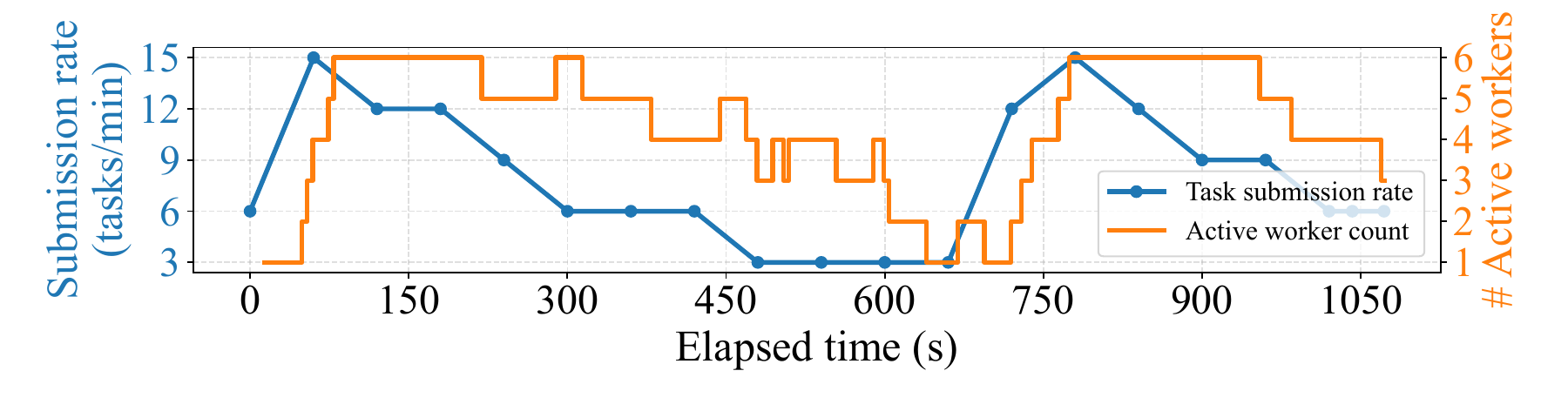}
    \vspace{-2em}
    \caption{Dynamic workload (new tasks/min) and the number of active workers in FlowMesh over time.}
    \label{fig:elastic-management}
\end{figure}

\textbf{Elasticity on Vast.ai.} To validate elasticity of FlowMesh, we conduct a dynamic-load experiment where the task arrival rate varies over time and track the number of active workers, using the experiment in \S\ref{exp1}. \autoref{fig:elastic-management} show the results, where the system effectively scales up and down when incoming workflows vary. A mild lag of 30-60 seconds is observed in elastic node creation and closure; it is attributed to the Vast.ai instance bidding API. Nonetheless, we consider this delay acceptable in practice.

\section{Related Work} \label{sec:relatedwork}
\noindent\textbf{LLM Serving Systems}
A number of recent systems focus on optimizing the \textit{inference} serving of large language models. {vLLM} \citep{kwon2023efficient} introduces a PagedAttention mechanism for dynamic memory management of the key-value cache, enabling efficient batching of autoregressive decoding. By reducing memory fragmentation and allowing cache reuse, vLLM achieves 2--4$\times$ higher throughput at similar latency compared to prior solutions like NVIDIA’s {FasterTransformer} \citep{nvidia2022fastertransformer} and {Orca} \citep{yu2022orca}. Orca is a distributed serving system that pioneered iteration-level scheduling and selective batching to better utilize GPU capacity during multi-token generation. Other frameworks such as {DeepSpeed-Inference} \citep{aminabadi2022deepspeed} apply optimized kernels and parallelism strategies to maximize throughput on hardware accelerators. These systems substantially improve runtime efficiency for deploying a trained LLM. However, their focus is on the \textit{inference phase} of a single model. In contrast, our work targets the \textit{post-training workflow} that produces and evaluates new model versions.

\vspace{0.05in}\noindent\textbf{Workflow Orchestration} Our work is related to general-purpose distributed execution frameworks that manage machine learning workflows. {Ray} \citep{moritz2018ray} supports a dynamic task graph and an actor-based abstraction to handle heterogeneous workloads. {Dask} \citep{rocklin2015dask} provides a dynamic scheduler for parallelizing Python computations, and {Modin} \citep{petersohn2020towards} scales data preprocessing via distributed DataFrame operations. {Alpa} \citep{zheng2022alpa} automates distributed parallelization of deep learning models. JellyBeanDS~\cite{wu2022serving} deploys complex ML workflows on a hybrid cloud with heterogeneous workers, optimizing for combined compute and networking costs. These frameworks are general and require users to manually stitch components of a complex LLM post-training pipeline. In contrast, our system provides an integrated infrastructure tailored for LLM post-training workflows.

\vspace{0.05in}\noindent\textbf{LLM Post-Training Workflows} After an LLM's initial pre-training, several techniques improve or adapt the model. {Supervised fine-tuning (SFT)} is used in FLAN and T0 \citep{wei2022finetuned, sanh2022multitask}, as well as in the initial stages of {InstructGPT} \citep{ouyang2022training}. However, SFT may not capture nuanced human preferences, motivating {RLHF} \citep{christiano2017deep}, which optimizes an LLM’s behavior using a reward model trained on human preferences. To reduce the human labeling burden, {RLAIF} \citep{lee2024rlaif} and {Constitutional AI} \citep{bai2022constitutional} propose using AI-generated feedback. Another approach is {agentic workflows}, where LLMs act as decision-makers. {ReAct} \citep{yao2023react} and {HuggingGPT} \citep{shen2023hugginggpt} represent such pipelines by combining reasoning and acting or orchestrating multiple models. While these works contribute to model improvement, our focus is on creating a shared infrastructure to support such workflows efficiently and at scale.

\section{Conclusion} 
\label{sec:conclusion}
In this work, we presented FlowMesh, a system that treats post-training and LLM-agent workloads as a shared, elastic service rather than isolated jobs. By unifying redundant work, batching compatible tasks, and scheduling them across diverse accelerators, FlowMesh improves efficiency and reduces cost, while its global control plane and stateless data plane enable scalable, portable, and resilient operation under real-world workloads.

\bibliographystyle{mlsys2025}
\bibliography{reference}

\appendix
 


\end{document}